# Electrically tunable radiative cooling performance of a photonic structure with thermal infrared applications

ATAOLLAH KALANTARI OSGOUEI,[1,2,][*] HASAN KOCER,[1] HALIL ISIK [1], YILMAZ DURNA,[1] AMIR GHOBADI, [1,3] AND EKMEL OZBAY [1,2,3,4, ][*]

[1]*NANOTAM-Nanotechnology Research Center, Bilkent University, 06800, Ankara, Turkey*
[2]*Department of Physics, Bilkent University, 06800, Ankara, Turkey*
[3]*Department of Electrical and Electronics Engineering, Bilkent University, 06800, Ankara, Turkey*
[4]*UNAM-Institute of Materials Science and Nanotechnology, Bilkent University, Ankara, Turkey*
*\*Corresponding author: akalantari@bilkent.edu.tr, ozbay@bilkent.edu.tr*

**Abstract**

**Thermal infrared (IR) radiation has attracted considerable attention due to its applications ranging from radiative cooling to thermal management. In this paper, we design a multi-band graphene-based metamaterial absorber compatible with infrared applications and radiative cooling performance. The proposed structure consists of the single-sized metal-insulator-metal (MIM) grating deposited on metal/insulator substrate and single-layer graphene. The system realizes a broadband perfect absorption ranging from 940 nm to 1498 nm and a narrowband perfect absorption at the resonance wavelength of 5800 nm. Meanwhile, the absorptivity of the structure is suppressed within the mid-wave infrared (MWIR) and long-wave infrared (LWIR) ranges. Furthermore, to demonstrate the tunability of the structure, an external voltage gate is applied to the single-layer graphene. It is shown that, by varying the chemical potential ($\mu_c$) of graphene layer from 0 eV to 1 eV, the absorption resonances at the mid-infrared (MIR) range can shift toward the shorter wavelengths. It is also observed that the structure can possess an average net cooling power over 18 W/m$^2$ at the ambient temperature, when $\mu_c$ is varied from 0 eV to 1 eV. Finally, we investigate the overall performances of the structure as a function of temperature to realize thermal infrared applications.**

## 1. Introduction

The ability to control the thermal infrared (IR) radiation is of great importance in a wide range of applications, including radiative cooling [1, 2], thermophotovoltaics (TPVs) [3, 4], and thermal infrared applications [5-10]. In particular, thermal radiation energy emitted from an object depends not only on the surface emissivity but also heavily on the fourth power of the absolute temperature according to the Stefan-Boltzmann equation $P = \varepsilon\sigma T^4$ (where $\sigma$ is the Steven-Boltzmann constant, and $\varepsilon$ and $T$ are the emissivity and absolute temperature of the surface, respectively) [11]. Therefore, reducing both the emissivity and the temperature of the surface are the two means to control the thermal radiation of the object. Temperature control represents a direct way of achieving thermal radiation, but it requires additional cooling and heating devices [12]. However, controlling the surface emissivity is more convenient than the temperature, as covering a low-emissivity material on the surface can efficiently suppress thermal radiation. Generally, when a low-emissivity material such as metal ($\varepsilon \approx 0.01$) is coated on a target surface, it is possible to reduce the emissivity of the coated object throughout the IR band [13]. By contrast, since emissivity is an intrinsic property of materials, the ability to tune the spectral emissivity is a difficult task. This is especially important when the IR radiation of the object is reduced, where the reduced emitting energy contributes to a sharp temperature increase in materials and, therefore, thermal instability may occur due to internal and external heat sources [14]. Therefore, multi-band thermal IR radiation with appropriate optical properties and reduced emittance through selective wavelengths are highly desirable. Metamaterial perfect absorbers (MPAs) represent a suitable candidate for achieving multi-band thermal IR radiation. In general, MPAs are artificially engineered materials for achieving near-perfect absorptions by utilizing a series of periodic arrayed unit cells such as plasmonic [15-19]. Through properly choosing geometric structures and materials, multi-band absorption peaks can be achieved at the specific wavelengths [20-22]. According to Kirchhoff's thermal radiation law, multi-band metamaterial absorber is equal to multi-band metamaterial emitter at the thermal equilibrium $\alpha(T, \lambda) = \varepsilon(T, \lambda)$ [23]. Thereby, metamaterial absorbers/emitters with wavelength-selectivity have a special ability to control thermal radiation. Recently, several plasmonic structures with metal-insulator-metal

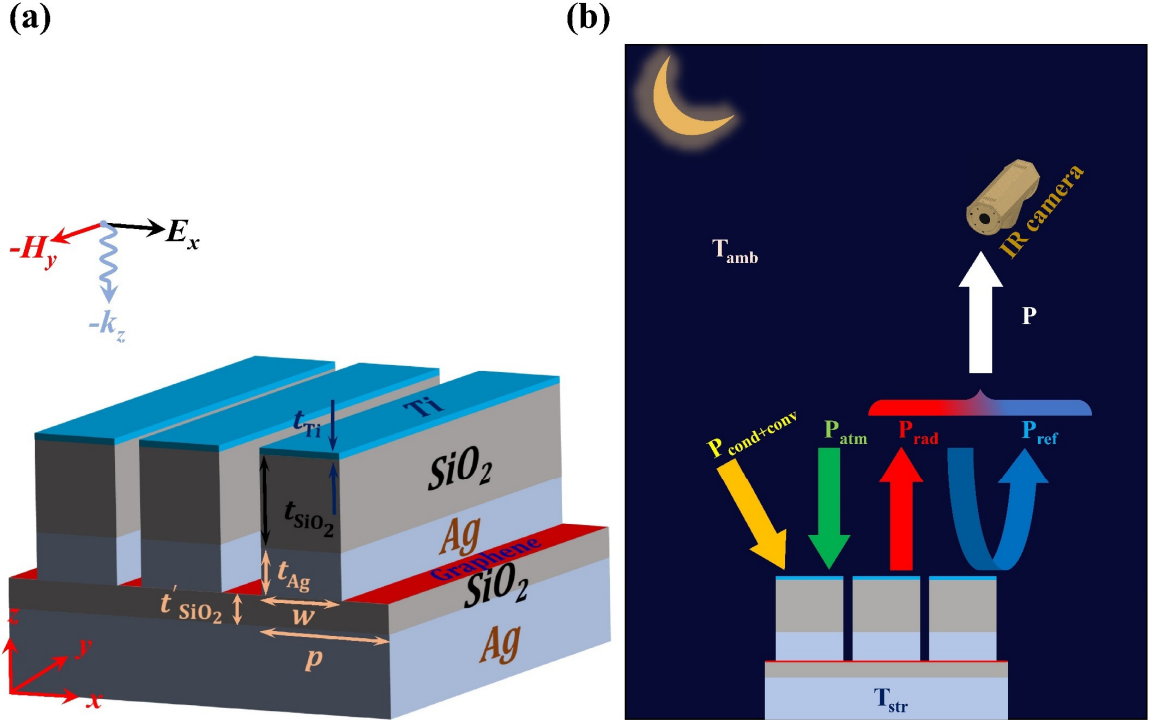

**Fig. 1.** (a) Schematic diagram of the proposed multi-band graphene-based metamaterial absorber/emitter. The proposed thermal infrared radiation structure consists of the single-sized MIM grating [an insulator layer of SiO2 ($t_{SiO_2} = 180$ nm) sandwiched between Ti ($t_{Ti} = 5$ nm) and the bottom Ag layer ($t_{Ag} = 70$ nm)] deposited on the metal/insulator ($t'_{SiO_2} = 30$ nm) substrate and the single-layer graphene. The unit cell is periodically arranged along the $x$ − direction with $p = 1200$ nm and the width of the MIM grating is $w = 1160$ nm. (b) Concept of the radiative cooling system. $P_{cond+conv}$ (yellow arrow) denotes the non-radiative heat transfer between the radiative cooling surface and the ambient, $P_{atm}$ (green arrow) is the total absorbed atmospheric radiation on the surface of the structure, $P_{rad}$ (red arrow) is the total thermal radiation emitted from the surface of the structure at $T_{str}$, and $P_{ref}$ (blue arrow) is the reflected environment radiation by the sample. The ambient temperature is assumed to be $T_{amb} = 20°C$ in the calculations.

(MIM) configurations [24, 25] and multi-layer structures [26, 27] have been used for the realization of thermal radiation. However, the resonant properties of the above-mentioned structures only present a static manipulation of the thermal radiations without flexible tunability control over the resonances, which greatly limits their extensive applications. Therefore, to realize dynamic control over the resonances, several phase-change materials (PCMs) like $Ge_2Sb_2Te_5$ (GST) and vanadium dioxide ($VO_2$) are proposed to control thermal radiation [28-30]. However, the low tunability and slow response speed of MPAs based on the GST and the $VO_2$ are the limiting factors to fully realize thermal radiation, due to the inherent ohmic losses of the materials [31, 32]. Therefore, graphene as a two-dimensional (2-D) material, which consists of a single layer of carbon atoms arranged in a lattice, has recently become one of the most attractive materials because of its interesting mechanically, chemically, and electrically tunable properties. Its high electron mobility and surface conductivity can be easily modulated by electrochemical potential by electronic or chemical doping. Due to the tunability of its electron mobility and conductivity, adjustable graphene-based metamaterial absorbers have been widely investigated in the IR and terahertz ranges [33-36]. Although the optical characteristics of the graphene has been extensively studied, the use of graphene-based metamaterial absorbers for dynamic control of thermal radiation has still remained unexplored because of its small absorption response ($\varepsilon < 0.02$) in the mid-infrared (MIR) region [37]. It is also feasible to design the few-layer graphene device to outperform the single-layer one in terms of the achievable tunability of the proposed structure. As several works have already been proposed to practically realize few-layer graphene structures for effectively tuning the resonances at the MIR region [38,

39]. Therefore, the few-layer graphene structure exhibits a similar trend as the single-layer one in terms of the tunability of the resonance wavelengths, and it is also observed that increasing the number of graphene layers can further increase the tunability of the resonances toward the shorter/longer wavelength spectrum.

Achieving multi-band thermal IR radiation technology compatible with radiative cooling is crucial to slow down the effects of global warming, which can mainly be utilized in building blocks and solar cells [40, 41]. For several decades, nighttime and daytime radiative cooling have been extensively studied in many research groups [42-45]. From the nighttime radiative cooling point view, radiators coated with inorganic thin films of silicon monoxide (SiO), silicon dioxide ($SiO_2$), silicon nitride ($Si_3N_4$), and other types of thin films have been reported [46-48]. With the recent technological progress in materials, a series of approaches for daytime radiative cooling have been proposed and developed to improve the device performances with high reflectivity for solar energy spectrum ($0.3 − 3$ μm) and a high IR thermal radiation within the atmospheric absorption window ($5 − 8$ μm). A traditional method to achieve daytime radiative cooling is to integrate a broadband thermal IR emission with solar reflectors to block the undesirable ranges from reaching the cooler by using a partially covered shield, such as polyethylene or ethylene [49, 50]. Other new approaches based on the photonic crystal and metamaterial structures have been developed for realizing daytime radiative cooling in recent years [51, 52]. However, the aforementioned existing radiative cooling structures are generally static in the sense that the resonance peaks are fixed without the possibility to tune the resonances. Therefore, there is a high demand to design dynamic radiative cooling since the external environment changes constantly in practice. To date, $VO_2$ and GST are widely used in the dynamic regulations of radiative properties due to their obvious phase change properties [29, 53, 54, 55]. In this paper, a multi-band graphene-based metamaterial absorber/emitter compatible with thermal IR applications and radiative cooling is proposed and numerically investigated using a commercial finite-difference time-domain (FDTD) software package solver [56]. The proposed structure consists of a single sized MIM grating embedded with a top highly lossy Titanium (Ti) and deposited on metal/insulator substrate. These two compact structures are separated by single-layer graphene. Conventional methods to fully achieve absorption are using ribbons, disks, and grating [57-61]. However, the fabrication of the graphene-based metamaterial involves time-consuming electron beam lithography and etching that make it a difficult task to scale up. Therefore, in this paper, we propose a graphene-based metamaterial emitter with a flat graphene sheet to overcome these difficulties for the future fabrication. Atomic layer lithography (ALD) is used as a new technology to fabricate sub-wavelength uniform features. In this process, ALD is implemented to fabricate nanogaps between the grating structure and deposited substrate films, providing an angstrom-scale lateral solution along the design. In this strategy, the upper nanograting pattern must be peeled off using the standard adhesive tape, which is an essential step for the ALD lithography step. Moreover, it is important to fabricate the vertical walls of the grating structure on the first layer to have a discontinuity between the other layers [62]. The proposed thermal IR system shows a broadband absorption spanning from 940 nm to 1498 nm (suitable for nighttime cooling), together with a narrowband resonant peak at the wavelength of 5800 nm that matches well with the atmospheric absorption window. Moreover, the absorptivity of the structure is suppressed within the mid-wave infrared (MWIR: $3 - 5$ μm) and the long-wave infrared (LWIR: $8 - 14$ μm) ranges representing the atmospheric transparency channels for electromagnetic (EM) waves. Meanwhile, it is shown that, by varying the external voltage applied to the single-layer graphene, the adjustability of the narrowband perfect absorption of the proposed structure at the MIR range can be well tuned to the shorter wavelengths. Theoretical and numerical analyses are also conducted to verify its cooling performances of the proposed electrically tunable metamaterial structure. The proposed design shows several distinctive advantages: (i) Multi-band IR and visual thermal infrared technology, (ii) efficient radiative cooling performances by radiation in the non-atmospheric window, (iii) simple design in fabrication due to the layers having the same size and shape, and (iv) tunable characteristics of the single-layer graphene by applying an external gate voltage, without the need to make new structures.

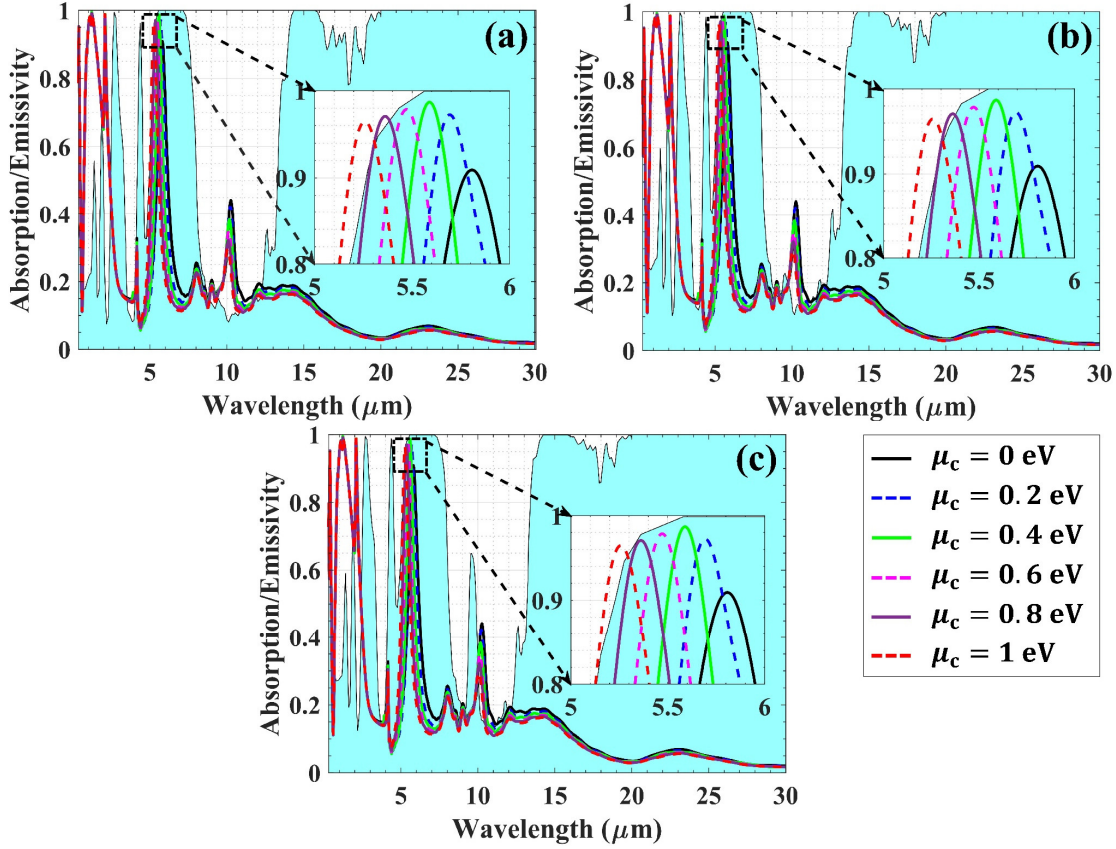

**Fig. 2.** Spectral absorptions of the proposed thermal infrared radiation system when an external voltage gate is applied to the single-layer graphene. Under different chemical potentials (0 eV ≤ $\mu_c$ ≤ 1 eV) at $T = 300$ K, the absorption resonance at the MIR range shift toward the shorter wavelengths. The blue shaded areas are the normalized atmospheric absorption spectrums, which are obtained by considering US standard 1976 atmospheric compositions at the vertical distances of (a) 2 km, (b) 5 km, and (c) 99 km.

## 2. Physical model and simulation

The schematic diagram of the proposed electrically tunable radiative cooling system compatible with thermal IR radiation is shown in Fig. 1(a). The unit cell consists of the single sized MIM grating (an insulator layer of $SiO_2$ sandwiched between the top highly lossy Ti

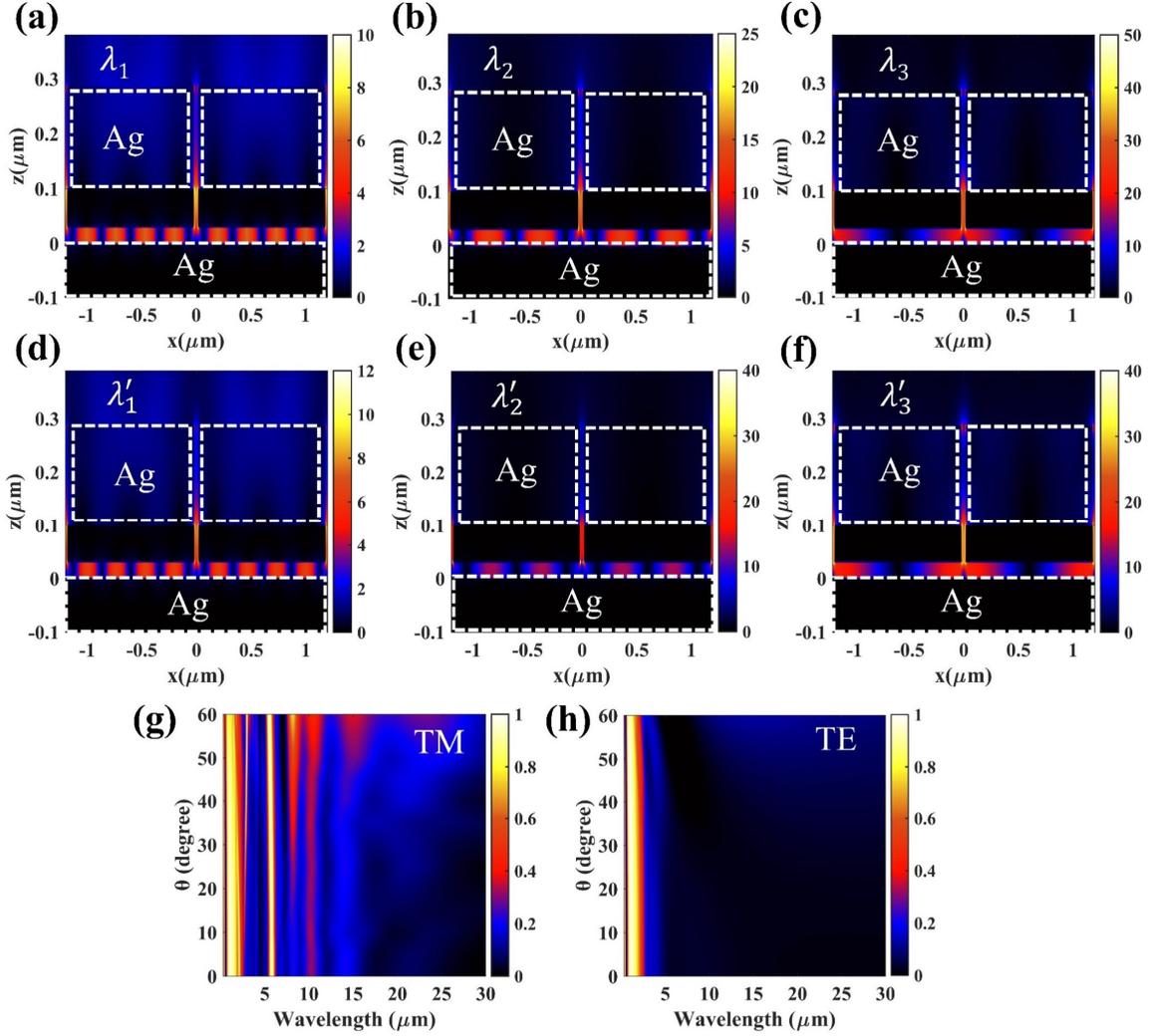

Fig.3. The total electric field distributions of the proposed multi-band graphene-based metamaterial absorber compatible with thermal infrared applications (without graphene layers) at the resonance wavelengths of (a) $\lambda_1 = 1230$ nm, (b) $\lambda_2 = 2110$ nm, and (c) $\lambda_3 = 5796$ nm, respectively. The total electric field distributions of the design (with the single-layer graphene when [$\mu_c = 0.6$ eV]) at the resonance wavelengths of (d) $\lambda'_1 = 1230$ nm, (e) $\lambda'_2 = 2110$ nm, and (f) $\lambda'_3 = 5519$ nm, respectively. The dependence of the absorption of the proposed design on the incident angle for (g) TM and (h) TE polarization modes.

and the metallic silver (Ag) layers) separated from the metal/insulator substrate and the single-layer graphene. Numerical simulations, based on the FDTD method with a commercial software package (Lumerical Solutions, Inc.), are utilized to optimize the geometrical parameters in such a way that a low emittance within the atmospheric transparency windows ($3 − 5$ μm, $8 − 14$ μm) and a high emittance in the atmospheric absorption window ($5 − 8$ μm) are achieved. Therefore, based on the simulation results, the optimized geometrical parameters are explicitly presented in the caption of Fig. 1(a). The thickness of each layer of the single-sized MIM grating structure is set as $t_{Ti} = 5$ nm, $t_{SiO_2} = 180$ nm, and $t_{Ag} = 70$ nm, respectively. The thickness of the insulator substrate is considered to be $t'_{SiO_2} = 30$ nm. The unit cell is periodically arranged along the $x$−direction with $p = 1200$ nm and the corresponding width of the MIM grating is $w = 1160$ nm. In the simulations, a uniform plane wave is normally propagating along the negative $−z$−direction with the polarization along the $x$−direction ($p$−polarization) and a wavelength range from $0.3 − 30$ μm is applied to investigate the optical properties of the proposed structure. Periodic boundary conditions are employed only along the $x$−direction and perfectly matching layers (PML) are applied along the $z$−direction to avoid the boundary scattering. It is important to mention here that the added single-sized MIM grating to the substrate can break polarization degeneracy. The proposed design is expected to act as thermal infrared applications in transverse magnetic (TM) mode, inversely has low absorption at transverse electric (TE) mode due to asymmetry through adding the single-sized grating layer. To overcome the polarization-dependent disadvantage in the structure, it is also possible to design periodical square plates. The reflectivity spectrum ($R$) is recorded by a 2D frequency-domain power monitor. Since the bottom Ag substrate is thicker than the penetration depth of the incident light, the transmission in our desired range is zero, and the absorption/emissivity of the design is calculated as $A = \varepsilon = 1 − R$, where $A$ and $R$ denote light absorption and reflection, respectively [63-65]. The frequency dependent relative permittivity of metallic Ag film is obtained from CRC Handbook of Chemistry and Physics [66] and the refractive index and

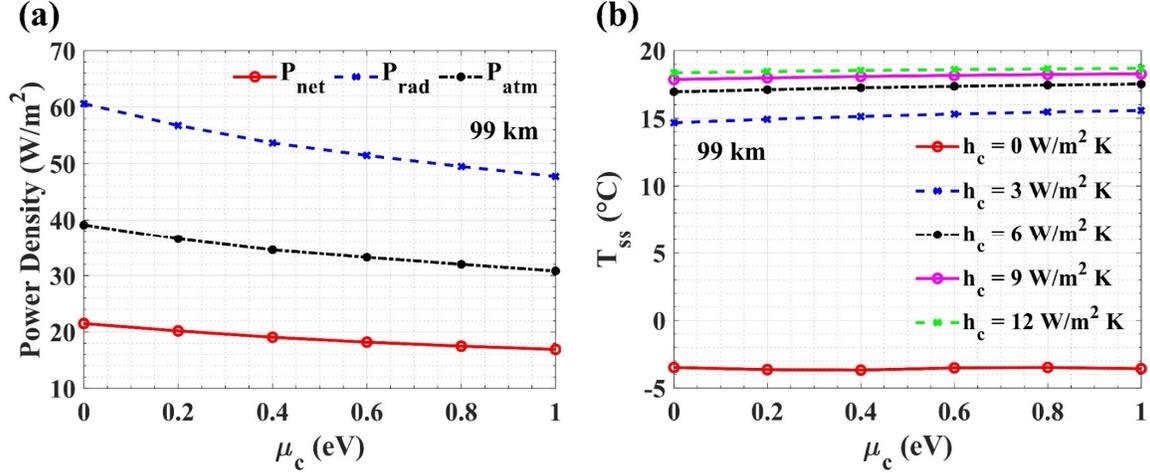

**Fig. 4.** (a) The net cooling power ($P_{net}$), absorbed atmospheric thermal radiation power ($P_{atm}$), and total thermal radiation power emitted from the surface ($P_{rad}$) of the proposed radiative cooling system as a function of $\mu_c$ varied from 0 eV to 1 eV by taking into consideration the atmospheric spectrum at the vertical distance of 99 km without considering the non-radiative heat transfer coefficient ($h_c = 0$ W/m²K). (b) The structure's steady-state temperature ($T_{ss}$) as a function of $\mu_c$ considering the effects of the non-radiative heat transfer coefficients ($h_c = 3, 6, 9, 12$ W/m²K). The ambient temperature is $T_{amb} = 20°$C in the calculations.

frequency-dependent SiO$_2$ Palik model are taken into account [67]. Moreover, following from the experimental data, complex dielectric constant of Ti is considered by the Palik model [67]. The single-layer graphene can be described as an infinitesimally thin, local two-sided surface characterized by the surface conductivity ($\sigma_s$). The surface conductivity of the single-layer graphene follows from the well-known Kubo equation, consisting of intraband ($\sigma_{intra}$) and interband terms ($\sigma_{inter}$) as follows [68]:

$$\sigma_s = \sigma_{intra} + \sigma_{inter}, \quad (1)$$

$$\sigma_{intra} = -j\frac{e^2 k_B T}{\pi \hbar^2 (\omega - j2\Gamma)}\left[\frac{\mu_c}{k_B T} + 2\ln\left(e^{-\mu_c/k_B T} + 1\right)\right], \quad (2)$$

$$\sigma_{inter} \simeq \frac{-je^2}{4\pi\hbar}\ln\left(\frac{2|\mu_c| - (\omega - j2\Gamma)\hbar}{2|\mu_c| + (\omega - j2\Gamma)\hbar}\right), \quad (3)$$

where $e$, $\omega$, $\hbar$, and $k_B$ are the charge of an electron, the angular-frequency of the plane wave, reduced Planck's and Boltzmann's constants, respectively. In addition, $\mu_c$ is the chemical potential, $\Gamma$ is the scattering rate, and $T$ is the temperature. In the simulation, we assume $T = 300$ K and $\Gamma = 0.0032$ eV. The surface conductivity of graphene sheet will be controlled by chemical potential ($\mu_c$) via applying electrostatic gating, which provides an effective way to tune the resonances of the proposed structure. One of the most appealing features of graphene is that its chemical potential can be readily adjusted across a large range by applying an external gate voltage (electrostatic biasing), resulting in a variety of surface conductivities. As a result, by adjusting the chemical potential via electrostatic biasing, it is feasible to modify the MIR resonance of the proposed structure. The relation between chemical potential level and the electrostatic biasing is given by an approximate closed-form expression as [69]:

$$\mu_c \approx \hbar v_F \sqrt{\frac{\pi C_d V_g}{e}}, \quad (4)$$

where $C_d = \varepsilon_d \varepsilon_0/t_s$ is the electrostatic gate capacitance, $t_s$ is the thickness of gate dielectric (bottom SiO$_2$ layer), $V_g$ is the applied gate voltage, and $v_F$ is the Fermi velocity ($1.0 \times 10^6$ m/s in graphene), respectively. As a result, the proposed structure's electrostatic biasing of the single-layer graphene may be done by connecting the graphene to the bottom Ag reflector (electrostatic ground) using conductive contacts and providing a gate voltage. The bias then enables independent controlling of the chemical potential of the graphene layer [62, 63]. To analyze the cooling performance, the net cooling power of the proposed structure is calculated by considering the temperature of the radiative cooler ($T_{str}$), and the ambient air temperature ($T_{amb}$). When the radiative cooling system is exposed to the night sky, it cannot be affected by the solar irradiance ($P_{sun} = 0$) and, therefore, the net cooling power ($P_{net}$) is significantly influenced by both the non-radiative heat transfer coefficient ($h_c$) and the atmospheric thermal radiation as shown in Fig. 1(b). Therefore, the net cooling power of the nighttime radiative cooler can be defined as:

$$P_{net}(T_{str}) = P_{rad}(T_{str}) - P_{atm}(T_{amb}) - P_{cond+conv}, \quad (5)$$

where $P_{rad}$ is the thermal radiation power emitted from the surface. $P_{atm}$ is the absorbed atmospheric radiation power on the surface of the radiative cooler, and $P_{cond+conv}$ denotes the non-radiative heat transfer, i.e., conduction and convection, between the radiative cooling surface and the ambient. Note that all of the power components above and below are considered to be power density (i.e., power divided by area). The total thermal radiation power per unit area ($P_{rad}$) from the surface of the radiative cooler is the rate at which radiation emitted at all possible wavelengths and directions, defined as:

$$P_{rad}(T_{str}) = \int \cos\theta d\Omega \int_0^\infty I_{bb}(\lambda, T_{str})\varepsilon_{str}(\Omega, \lambda)d\lambda, \quad (6)$$

where $\int d\Omega = 2\pi \int_0^{\pi/2} d\theta \sin\theta$ is the angular integral over a hemisphere. $I_{bb}(\lambda, T_{str}) = \frac{2hc^2}{\lambda^5 [\exp(hc/\lambda k_B T_{str}) - 1]}$ is the spectral radiance of a blackbody at temperature $T_{str}$, where $c$ is the speed of light in vacuum, $\lambda$ is the wavelength, and $\varepsilon_{str}(\Omega, \lambda)$ is the directional emissivity of the structure as a function of wavelength. The total absorbed power due to the incident atmospheric thermal radiation on the radiative cooling surface can also be expressed as:

$$P_{atm}(T_{amb}) = \int \cos\theta d\Omega \int_0^\infty I_{bb}(\lambda, T_{amb})\varepsilon_{atm}(\Omega, \lambda)\varepsilon_{str}(\Omega, \lambda)d\lambda, \quad (7)$$

here $\varepsilon_{atm}(\Omega, \lambda) = 1 - t(\lambda)^{1/\cos\theta}$ is the angle-dependent emissivity of the atmosphere as a function of wavelength, where $t(\lambda)$ is the atmospheric transmittance in the zenith direction, and $I_{bb}(\lambda, T_{amb})$ is the blackbody spectral radiance at temperature $T_{amb}$, defined as $I_{bb}(\lambda, T_{amb}) = \frac{2hc^2}{\lambda^5[\exp(hc/\lambda k_B T_{amb})-1]}$. In addition to the thermal radiation power, the radiative cooling surface is generally subject to other non-radiative heat transfer with the surrounding environment [64]. Therefore, the power loss due to the convection and conduction heat transfer can be expressed as:

$$P_{cond+conv} = h_c(T_{amb} - T_{str}), \qquad (8)$$

where $h_c$ denotes the combined non-radiative heat transfer coefficient that accounts for the effect of conductive and convective heat transfer between the surface of the radiative cooler and the external surface.

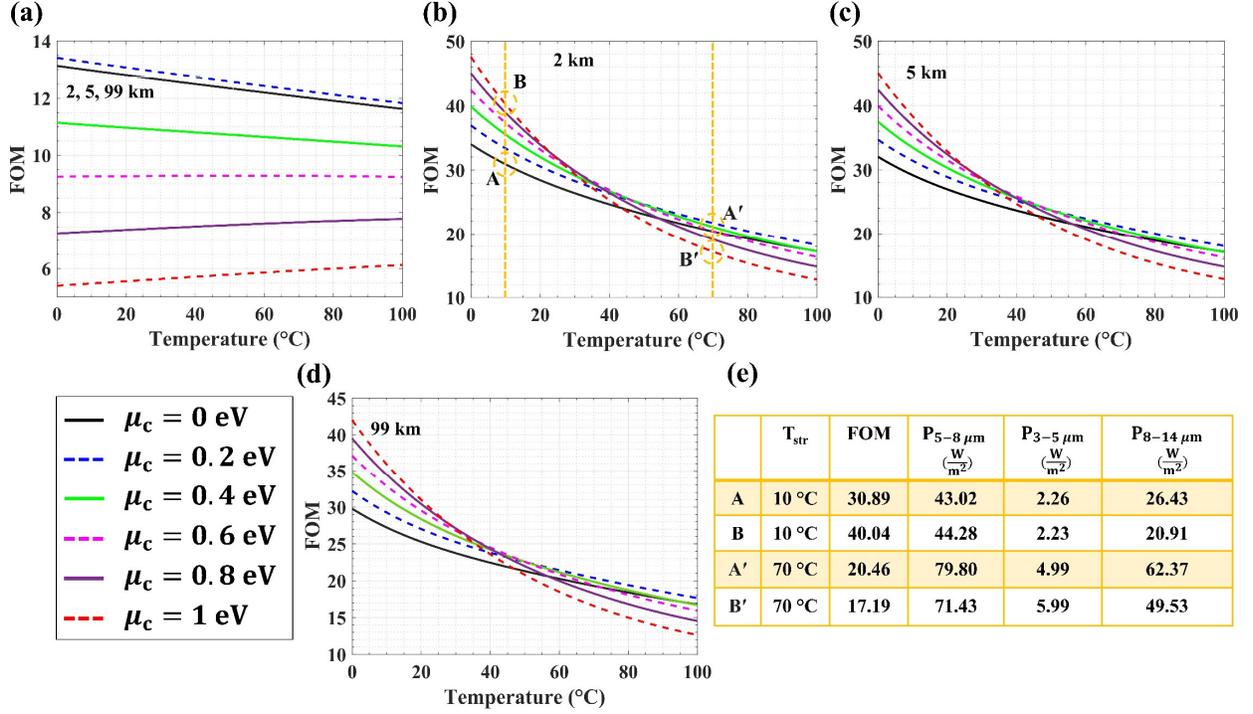

**Fig. 5.** (a) The overall performances (FOM) of the proposed radiative cooling system without considering the effects of the reflected ambient radiation power ($P_{ref} = 0$) as a function of the structure's temperature ($0°C \leq T_{str} \leq 100°C$) when the $\mu_c$ is varied from $0\ eV$ to $1\ eV$ for different atmospheric absorption spectra. The overall performances (FOM) of the proposed structure contains the sample radiation power ($P_{rad}$) and the reflected ambient radiation power ($P_{ref}$) as a function of structure's temperature ($0°C \leq T_{str} \leq 100°C$) when the $\mu_c$ is varied from $0\ eV$ to $1\ eV$ by considering the atmospheric compositions at the vertical distances of (b) 2 $km$, (c) 5 $km$, and (d) 99 $km$. (e) The total radiation power detected at the working wavelength bands of MWIR, LWIR, and atmospheric absorption window, along with FOM values at different structure's temperature of $T_{str} = 10°C$ and $T_{str} = 70°C$.

## 3. Results and discussions

The simulated results of the emissivity/absorptivity of the proposed radiative cooling system compatible with thermal IR system in the wavelength range of $0.3 - 30$ µm are shown explicitly in Figs. 2(a)-2(c). The blue areas are the normalized atmospheric absorption bands, which are obtained using a commercial software package (MODTRAN) by considering US standard 1976 atmospheric compositions at different vertical distances (altitudes) of 2 km, 5 km, and 99 km through the atmosphere from the ground level [72]. It is observed in the figures that the proposed system exhibits a high absorption with over 90% in a wide wavelength range from 940 nm to 1498 nm (suitable for nighttime radiative cooling) and a narrowband perfect absorption at the wavelength of 5800 nm (within the atmospheric absorption window) is sufficiently excited when $\mu_c = 0$ eV. Meanwhile, the proposed structure shows remarkable low absorptivity/emissivity within the MWIR and LWIR ranges, corresponding to the atmospheric transparency windows. The results also reveal that an undesired absorption peak with over 40% absorptivity is excited in the LWIR range because of the intrinsic vibrational modes (optical phonons) of the $SiO_2$ layer [73]. Moreover, the thermos-optics coefficients of optical materials (except phase change materials) are in the range of $10^{-4}$. Therefore, the obtained absorption spectra of the proposed design are independent of the temperature variations [74]. To demonstrate the tunability performances of the proposed graphene-based IR thermal infrared system, the simulated absorption spectra as a function of $\mu_c$ varied from 0 eV to 1 eV are also shown in Figs. 2(a)-2(c). It should be important to mention that by applying an external gate voltage or means of chemical doping, the chemical potential ($\mu_c$), and thus the surface conductivity ($\sigma_s$) of the single-layer graphene can be easily controlled. When the chemical potential of the single layer graphene is increased toward 1 eV, the absorption resonances at the MIR range can shift toward the shorter wavelengths, while the obtained broadband response at the near-infrared (NIR) region stay almost unchanged, due to the single-layer graphene characteristics [69]. In addition, it is clear that, by increasing the chemical potential ($\mu_c$), the amount of absorption peak at the LWIR range, caused by the vibrational modes of $SiO_2$ layer, decreases gradually. This can be deduced from the fact that a relatively higher amount of incident light is reflected from the surface of graphene as the surface conductivity ($\sigma_s$) of the single-layer graphene increases. It is also shown that the absorption peak (intensity) at the resonance wavelength of 5800 nm increases when the chemical potentials ($\mu_c$) changes from 0 eV to 0.2 eV, and stays almost constant for $\mu_c > 0.2$ eV, as shown in Figs.

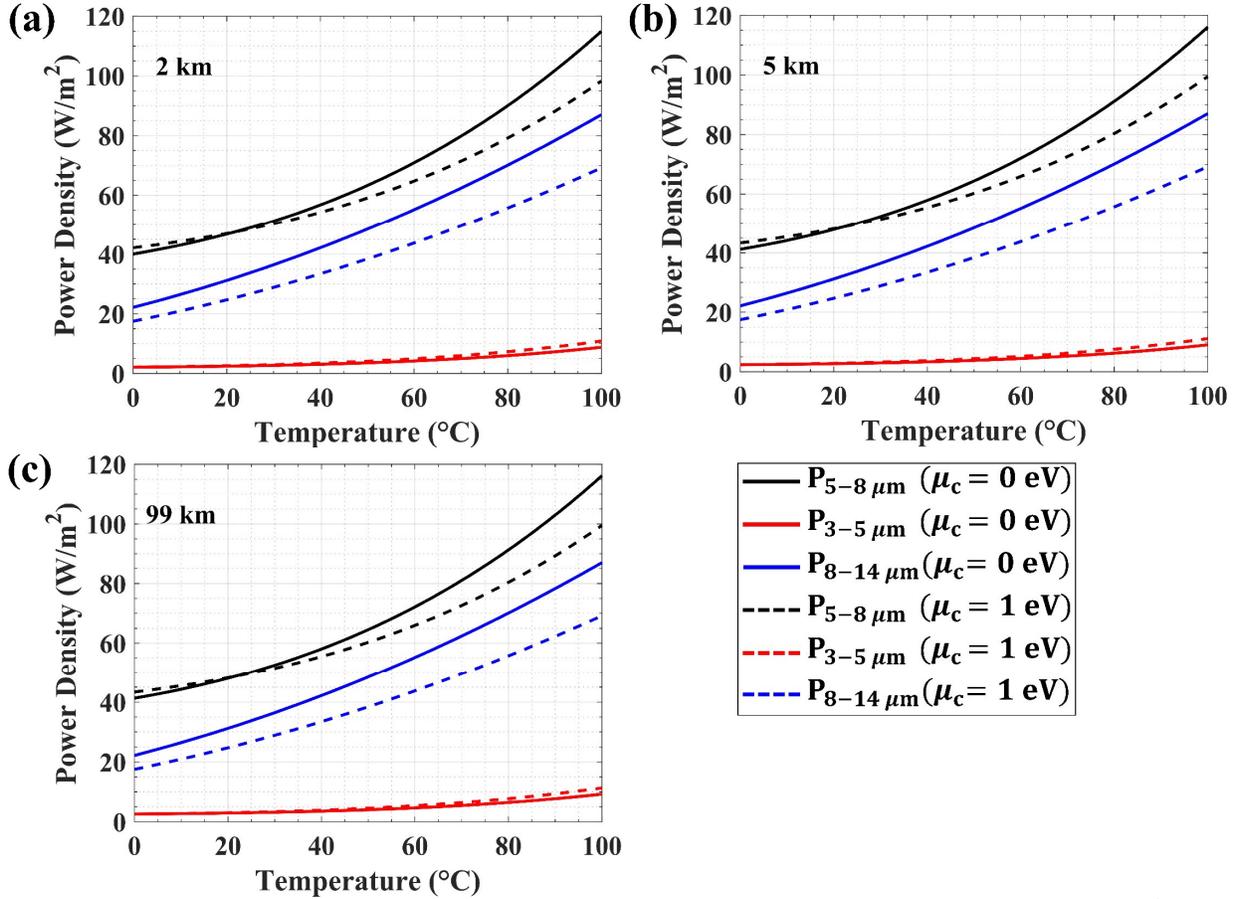

**Fig. 6.** The total radiation power of the proposed structure in the LWIR, MWIR, and atmospheric absorption bands as a function of structure's temperature ($0°C \leq T_{str} \leq 100°C$) when $\mu_c = 0, 1$ eV and by considering the atmospheric compositions at the vertical distances (altitude) of (a) 2 km, (b) 5 km, and (c) 99 km.

2(a)-2(c). This is due to the fact that the impedance matching improves by increasing the chemical potential, according to the effective-medium theory [75]. All in all, the proposed electrically tunable graphene-based metamaterial emitter absorption spectra perfectly match the atmospheric absorption window, while maintaining low emissivity within the atmospheric transparency bands of ($3 - 5$ μm, $8 - 14$ μm) with adjustability characteristics that make it suitable for nighttime radiative cooling and thermal infrared technology due to the wideband absorption within a spectral range of the NIR region.

## 4. Physical Mechanism and angular dependence

In order to better understand the physical mechanism of the proposed design under the normal incidence, the electric field distributions at the resonance wavelengths are obtained and illustrated in Fig. 3, corresponding to the cases without graphene and with the single layer graphene ($\mu_c = 0.6$ eV). It is observed from Figs. 3(a) and 3(d) at the resonance wavelengths of $\lambda_1 = \lambda'_1 = 1230$ nm, the incident light penetrates through the thin Ti layer and reflects back at the bottom metallic layer. The two reflectors clearly excite the Fabry-Perot (FP) cavity resonance leading to a broaden absorption response due to the lossy behaviour of Ti in the NIR region [76]. On the other hands, the electric-field distributions at the other excited resonances are different. In particular, the electric field at the resonance wavelengths of $\lambda_2 = \lambda'_2 = 2110$ nm are mostly localized inside the spacer layer due to the excitation of the gap surface plasmon resonances (GSPs) with the third-order mode, while as it can be seen in Figs. 3(c) and 3(f), the absorptions at the resonance wavelengths of $\lambda_3 = 5796$ nm and $\lambda'_3 = 5519$ nm are attributed to the excitation of GSPs with the first-order mode [77-79]. The absorptions of the proposed multi-band graphene-based metamaterial absorber compatible with thermal infrared applications for both TM and TE polarized modes are also obtained and shown in Figs. 3(g) and (h). It is observed from the Fig. 3(g) that the nearly perfect broadband and narrowband resonances of the proposed design are preserved up to 60 degrees of the incident angle for TM polarization modes, while another narrowband resonance is excited at the longer wavelengths when the incident angle is 50 degrees. For TE polarization mode as shown in Fig. 3(h), there is only a single broadband absorption peak excited and preserved up to 60 degrees of the incident light.

## 5. Radiative cooling performances

In order to better understand the efficiency analysis of the proposed electrically tunable graphene-based radiative cooler, the cooling performance of the system is investigated by calculating $P_{rad}$ and $P_{atm}$ at wavelengths with a range of $0.3 - 30$ μm and by taking into consideration the atmospheric absorption spectrum at the vertical distance of 99 km. The ambient temperature is assumed to be $T_{amb} = 20°C$ in the calculations. Figure 4(a) shows the net cooling power ($P_{net}$), absorbed atmospheric thermal radiation power ($P_{atm}$), and total thermal radiation power ($P_{rad}$) emitted from the surface of the cooler as a function of $\mu_c$ varied from 0 eV to 1 eV. It can be clearly seen that the proposed cooler at nighttime (no solar irradiance, $P_{sun} = 0$) achieves a net cooling power over 21.5 W/m$^2$ at ambient temperature, when $\mu_c = 0$ eV. As the $\mu_c$ increases from 0 eV to 1 eV, as shown in Fig. 4(a), the net cooling power can be reduced to

16.9 W/m² without considering the non-radiative heat transfer coefficient ($h_c = 0$ W/m²K). In this way, the proposed electrically tunable structure can meet the requirement of the nighttime radiative cooling, achieving the average net cooling power of over 18 W/m² by varying $\mu_c$ from 0 eV to 1 eV. Figure 4(b) shows the structure's steady-state temperature ($T_{ss}$) as a function of $\mu_c$ with and without considering the effects of the non-radiative heat transfer coefficient ($h_c$). Note that the steady-state temperature is obtained by solving eq. (4) with $P_{net}(T_{str} = T_{ss}) = 0$. It is clear that in the absence of the non-radiative heat transfer a reduction of 23.4°C below the ambient temperature (20°C) under clear night sky is achieved at $\mu_c = 0$ eV. When the chemical potential of the single-layer graphene continues to rise (0.2 eV $\leq \mu_c \leq$ 1 eV), the proposed structure's steady-state temperature gradually decreases, achieving a maximum temperature decrease of 23.6°C, when the structure is tuned at $\mu_c = 0.4$ eV. Moreover, to quantify non-radiative heat transfer coefficients on the overall temperature reduction of the radiative cooling surface at steady-state, we need to accurately determine the overall effects of non-radiative heat transfer coefficient $h_c$, when the radiative cooling surfaces are exposed to different surrounding environments. In general, four different non-radiative heat transfer coefficients, namely $h_c = 3, 6, 9, 12$ W/m²K are considered for the practical applications of the radiative cooling. As can be seen in Fig. 4(b), the average temperature decreases of the proposed radiative cooler are 4.8°C ($h_c = 3$ W/m²K), 2.7°C ($h_c = 6$ W/m²K), 1.9°C ($h_c = 9$ W/m²K), and 1.6°C ($h_c = 12$ W/m²K) below the ambient temperature. It can be concluded from these results that the cooling rate of the system decreases when the non-radiative heat transfer coefficient is increased. Therefore, the proposed radiative cooling system shows a remarkable nighttime cooling performance under different non-radiative heat transfer coefficients. In order to characterize the overall performance of the proposed system, the figure of merit (FOM) is defined as follows [80]:

$$\text{FOM} = \frac{P_{5-8\,\mu m}^2}{P_{MWIR} \cdot P_{LWIR}}, \quad (9)$$

where $P_{5-8\,\mu m}$, $P_{MWIR}$, and $P_{LWIR}$ represent the radiated powers emitted from the surface of the structure in the atmospheric absorption, MWIR, and LWIR spectral windows, respectively. A high value of FOM can be obtained from the structure with low emissivity within the atmospheric transparency windows (MWIR and LWIR ranges), and high emissivity within the atmospheric absorption band. The FOM values of the proposed radiative cooling system are evaluated according to the total amount of power emitted from the surface of the object. The total power detected from the IR camera as shown in Fig. 1(b), contains the self-radiation power ($P_{rad}$) and the reflected ambient radiation power by the object ($P_{ref}$) as [80, 81]:

$$P(\varepsilon_{str}, \varepsilon_{atm}, T_{str}, T_{amb}) = P_{rad}(\varepsilon_{str}, T_{str}) + P_{ref}(\varepsilon_{str}, \varepsilon_{atm}, T_{str})$$
$$= \int_{\lambda_2}^{\lambda_1} \varepsilon_{str}(\lambda) \cdot \frac{2\pi hc^2}{\lambda^5} (e^{\frac{hc}{\lambda k_B T_{str}}} - 1)^{-1} d\lambda + \int_{\lambda_2}^{\lambda_1} [1 - \varepsilon_{str}(\lambda)] \cdot \varepsilon_{atm}(\lambda) \cdot \frac{2\pi hc^2}{\lambda^5} (e^{\frac{hc}{\lambda k_B T_{atm}}} - 1)^{-1} d\lambda,$$
(10)

where $\lambda_1$ and $\lambda_2$ represent the working wavelengths of the IR camera. The FOM values of the proposed multi-band radiative cooling system *without considering the effects of the reflected ambient radiation power* ($P_{ref} = 0$) as a function of the structure's temperature (0°C $\leq T_{str} \leq$ 100°C), when $\mu_c$ is varied from 0 eV to 1 eV is shown in Fig. 5(a). It is clear that FOM values of the proposed cooling system do not change, under different atmospheric emissivity profiles since the reflected power is not included and the sample radiation power is independent of the atmospheric emissivity. Moreover, as shown in Fig. 5(a), the proposed structure shows relatively lower FOM when $\mu_c = 0$ eV, than when it is tuned to $\mu_c = 0.2$ eV. This is deduced from the fact that the impedance matching condition is actively achieved by increasing $\mu_c$ [75]. In addition, as the temperature increases, the performances of the proposed structure (FOM) decrease when $\mu_c = 0$ eV, 0.2 eV, and 0.4 eV because the radiation energy in the MWIR and LWIR bands increases. On the other hand, when $\mu_c$ of the single-layer graphene is tuned to 0.8 eV, and 1 eV, the overall performances of the proposed cooling system increase, as the temperature increases. However, for $\mu_c = 0.6$ eV it is seen that the performance remains nearly constant while increasing the temperature. We next demonstrate *the total radiated and reflected power on the overall performances* of the proposed structure. Figures 5(b) -5(d) show the overall performances of the proposed nighttime radiative cooler obtained from the simulated results as a function of structures' temperature (0°C $\leq T_{str} \leq$ 100°C), when $\mu_c$ is varied from 0 eV to 1 eV. It is clear from the figures that the overall performances (FOM) exhibit similar trends under different altitude (2 km, 5 km, and 99 km) atmospheric emissivity such that as the temperature increases, the performances (FOM) of the structure decrease. Moreover, the overall performances (FOM) of the radiative cooling system decrease with higher altitude (i.e., 99 km) atmospheric emissivity, since the reflected ambient radiation power ($P_{ref}$) is related to the atmospheric absorption spectrum. It is also evident from Figs. 5(b)-5(d) that, as the temperature increases, the proposed structure exhibits a sharper reduction of the FOM values when $\mu_c$ is increasing. The total radiation power detected by the IR camera at the working wavelength bands of MWIR ($P_{3-5\,\mu m}$), LWIR ($P_{8-14\,\mu m}$), and atmospheric absorption window ($P_{5-8\,\mu m}$), together with FOM values at different structure's temperature of $T_{str} = 10$°C, and $T_{str} = 70$°C are all summarized in Fig 5(e). To further demonstrate how the overall performances (FOM) of the proposed radiative cooling system behaves as the temperature increases, we consider two different $\mu_c$ cases, when $\mu_c = 0$ eV, and $\mu_c = 1$ eV. Figures 6(a)-6(c) show the total radiation power in the MWIR, LWIR, and atmospheric absorption bands for different altitude atmospheric emissivities. The results indicate that the proposed structure has higher total radiation powers in the LWIR ($P_{8-14\,\mu m}$), and atmpohseric absorption window ($P_{5-8\,\mu m}$), as the temperature increases. Moreover, as can be seen in Fig. 6(a), the proposed structure exhibits a lower radiation power in the LWIR band ($P_{8-14\,\mu m} = -20.5\%$) compared to the radiation power in the atmospheric window ($P_{5-8\,\mu m} = -14.5\%$), when $\mu_c$ is varied from 0 eV to 1 eV at $T_{str} = 100$°C. However, the radiation power in the MWIR band ($P_{3-5\,\mu m}$) is almost independent of the temperature variations. That is the main reason why the overall performance of the proposed structure shows a lower value as the temperature and $\mu_c$ increase. Clearly, due to the high FOM values, the proposed structure shows a promising way to realize the nighttime radiative performances with thermal IR radiation technology.

## 5. Conclusion

In conclusion, we have proposed a multi-band graphene-based metamaterial absorber/emitter compatible with nighttime radiative cooling and thermal IR applications. The proposed structure consists of the single-sized MIM grating (an insulator layer of SiO₂ sandwiched between the top highly lossy Ti and Ag layers) deposited on the metal/insulator substrate and the single- layer graphene. Numerical simulations based on an FDTD solver are utilized to investigate the optical properties of the proposed design. It is observed that a broadband perfect absorption spanning from 940 nm to 1498 nm (suitable for nighttime radiative cooling) and a narrowband perfect absorption at the wavelength of 5800 nm (within the atmpohseric absorption window) are successfully achieved. Meanwhile, the proposed IR system shows relatively low emissivity within the atmospheric transparency windows $(3-5\,\mu m, 8-14\,\mu m)$. Furthermore, the tunability characteristics of the proposed design are demonstrated when an external voltage gate is applied to the single-layer graphene. The proposed design was also optimized using practical geometrical parameters. As a result, the numerical results reported in this paper can be fabricated and verified. Next, the cooling efficiency of the proposed design is validated by calculating $P_{\text{rad}}$ and $P_{\text{atm}}$ in the wavelength range of $0.3-30\,\mu m$. From the simulation-based analysis of the cooling ability, the proposed electrically tunable structure can achieve the average net cooling power of over $18\,W/m^2$, when $\mu_c$ is varied from $0\,eV$ to $1\,eV$. Finally, we investigate the overall performances (FOM) of the proposed structure as a function of temperature. Due to the high FOM values, the proposed structure achieves a promising way to simultaneously realize the nighttime radiative cooling and thermal IR applications. This work sheds light on multispectral and adjustable thermal management with IR performance and thermal infrared applications.

**Acknowledgment**. E. Ozbay acknowledges partial support from the Turkish Academy of Sciences (TUBA).

**Authors' Contributions.** First author (A.K.O.) carried out the modeling, design, and simulations. H.K and H.I. and Y.D. and A.G. assisted in theoretical review and simulations. E.O. supervised the study. All the authors contributed in the results, discussions, and paper writing.

**Disclosures**. The authors declare no conflicts of interest.

## References

1. Raman, A. P., Abou Anoma, M., Zhu, L., Rephaeli, E., & Fan, S. (2014). Passive radiative cooling below ambient air temperature under direct sunlight. *Nature*, *515*(7528), 540-544.
2. Li, T., Zhai, Y., He, S., Gan, W., Wei, Z., Heidarinejad, M., ... & Hu, L. (2019). A radiative cooling structural material. *Science*, *364*(6442), 760-763.
3. Lenert, A., Bierman, D. M., Nam, Y., Chan, W. R., Celanović, I., Soljačić, M., & Wang, E. N. (2014). A nanophotonic solar thermophotovoltaic device. *Nature nanotechnology*, *9*(2), 126-130.
4. Shemelya, C., DeMeo, D., Latham, N. P., Wu, X., Bingham, C., Padilla, W., & Vandervelde, T. E. (2014). Stable high temperature metamaterial emitters for thermophotovoltaic applications. *Applied Physics Letters*, *104*(20), 201113.
5. Osgouei, A. K., Khalichi, B., Omam, Z. R., Ghobadi, A., & Özbay, E. (2022, September). A Wavelength-Selective Multilayer Absorber for Heat Signature Control. In *2022 International Conference on Electromagnetics in Advanced Applications (ICEAA)* (pp. 190-191). IEEE.
6. Khalichi, B., Omam, Z. R., Osgouei, A. K., Ghobadi, A., & Özbay, E. (2022, July). A Transmissive All-Dielectric Metasurface-Based Nanoantenna Array for Selectively Manipulation of Thermal Radiation. In *2022 IEEE International Symposium on Antennas and Propagation and USNC-URSI Radio Science Meeting (AP-S/URSI)* (pp. 21-22). IEEE.
7. Khalichi, B., Omam, Z. R., Osgouei, A. K., Ghobadi, A., & Özbay, E. (2022, July). Polarization Insensitive Phase Change Material-Based Nanoantenna Array for Thermally Tunable Infrared Applications. In 2022 IEEE International Symposium on Antennas and Propagation and USNC-URSI Radio Science Meeting (AP-S/URSI) (pp. 687-688). IEEE.
8. Osgouei, A. K., Ghobadi, A., Khalichi, B., & Ozbay, E. (2021). A spectrally selective gap surface-plasmon-based nanoantenna emitter compatible with multiple thermal infrared applications. Journal of Optics, 23(8), 085001.
9. Osgouei, A. K., Buhara, E., Khalichi, B., Ghobadi, A., & Ozbay, E. (2021, October). Wavelength selectivity in a polarization-insensitive metamaterial-based absorber consistent with atmospheric absorption windows. In 2021 IEEE Photonics Conference (IPC) (pp. 1-2). IEEE.
10. Thermally Tunable from Narrowband to Broadband Metamaterial-Based Nanoantenna Emitter
11. Catalanotti, S., Cuomo, V., Piro, G., Ruggi, D., Silvestrini, V., & Troise, G. (1975). The radiative cooling of selective surfaces. *Solar Energy*, *17*(2), 83-89.
12. Kim, H., Kim, K., & Kim, W. (2017). Research for Actively Reducing Infrared Radiation by Thermoelectric Refrigerator. *Transactions of the Korean Society of Mechanical Engineers B*, *41*(3), 199-204.
13. Hao, T., Hai-Tao, L., & Hai-Feng, C. (2013). A thin radar-infrared stealth-compatible structure: Design, fabrication, and characterization. *Chinese Physics B*, *23*(2), 025201.
14. Yunus A., Çengel, & Ghajar, A. J. (2015). *Heat and Mass Transfer: Fundamentals & Applications*. McGraw Hill Education.
15. Xiao, Z., Lv, F., Li, W., Zou, H., & Li, C. (2021). A three-dimensional ultra-broadband and polarization insensitive metamaterial absorber and application for electromagnetic energy harvesting. *Waves in Random and Complex Media*, *31*(6), 2168-2176.
16. Osgouei, A. K., Hajian, H., Khalichi, B., Serebryannikov, A. E., Ghobadi, A., & Ozbay, E. (2021). Active Tuning from Narrowband to Broadband Absorbers Using a Sub-wavelength VO$_2$ Embedded Layer. *Plasmonics*, 1-9.
17. Mulla, B., & Sabah, C. (2015). Perfect metamaterial absorber design for solar cell applications. *Waves in Random and Complex Media*, *25*(3), 382-392.
18. Khalichi, B., Ghobadi, A., Osgouei, A. K., & Ozbay, E. (2021). Diode like high-contrast asymmetric transmission of linearly polarized waves based on plasmon-tunneling effect coupling to electromagnetic radiation modes. *Journal of Physics D: Applied Physics*.
19. Osgouei, A. K., Ghobadi, A., Khalichi, B., Sabet, R. A., Tokel, O., & Ozbay, E. (2022). Visible light metasurface for adaptive photodetection. *Journal of Physics D: Applied Physics*, *55*(47), 475103.
20. Osgouei, A. K., Hajian, H., Serebryannikov, A. E., & Ozbay, E. (2021). Hybrid indium tin oxide-Au metamaterial as a multiband bi-functional light absorber in the visible and near-infrared ranges. *Journal of Physics D: Applied Physics*, *54*(27), 275102.
21. Buhara, E., Ghobadi, A., Khalichi, B., Kocer, H., & Ozbay, E. (2021). Mid-infrared adaptive thermal camouflage using a phase-change material coupled dielectric nanoantenna. *Journal of Physics D: Applied Physics*, *54*(26), 265105.
22. Osgouei, A. K., Khalichi, B., Buhara, E., Ghobadi, A., & Ozbay, E. (2021, October). Dual-band polarization insensitive metamaterial-based absorber suitable for sensing applications. In 2021 IEEE Photonics Conference (IPC) (pp. 1-2). IEEE.
23. Agassi, J. (1967). The kirchhoff-planck radiation law. *Science*, *156*(3771), 30-37.
24. Khalichi, B., Omam, Z. R., Osgouei, A. K., Ghobadi, A., & Özbay, E. (2022, July). Highly One-Way Electromagnetic Wave Transmission Based on Outcoupling of Surface Plasmon Polaritons to Radiation Modes. In 2022 IEEE International Symposium on Antennas and Propagation and USNC-URSI Radio Science Meeting (AP-S/URSI) (pp. 912-913). IEEE.
25. Omam, Z. R., Khalichi, B., Osgouei, A. K., Ghobadi, A., & Özbay, E. (2022, September). Multi-Band Light-Matter Interaction in hBN-Based Metasurface Absorber. In 2022 International Conference on Electromagnetics in Advanced Applications (ICEAA) (pp. 091-092). IEEE.

26. Osgouei, A. K., Ghobadi, A., Khalichi, B., Sabet, R. A., Tokel, O., & Ozbay, E. (2022). Visible light metasurface for adaptive photodetection. Journal of Physics D: Applied Physics, 55(47), 475103.
27. Peng, L., Liu, D., Cheng, H., Zhou, S., & Zu, M. (2018). A multilayer film based selective thermal emitter for infrared stealth technology. *Advanced Optical Materials*, *6*(23), 1801006.
28. Ye, H. N., Zhang, X. L., Zhao, Y., & Zhang, H. F. (2021). A tunable metasurface based on Vanadium dioxide for Broadband RCS reduction. *Waves in Random and Complex Media*, 1-12.
29. Osgouei, A. K. (2021). Plasmonic Metamaterial Based Structures for Designing of Multiband and Thermally Tunable Light Absorbers, Multiple Thermal Infrared Emitter, and High-contrast Asymmetric Transmission Optical Diode (Doctoral dissertation, Bilkent Universitesi (Turkey)).
30. Cao, T., Wang, R., Simpson, R. E., & Li, G. (2020). Photonic Ge-Sb-Te phase change metamaterials and their applications. *Progress in Quantum Electronics*, 100299.
31. Hutchins, M. G., Butt, N. S., Topping, A. J., Gallego, J., Milne, P., Jeffrey, D., & Brotherston, I. (2001). Infrared reflectance modulation in tungsten oxide based electrochromic devices. *Electrochimica acta*, *46*(13-14), 1983-1988.
32. Huang, Y., Boriskina, S. V., & Chen, G. (2014). Electrically tunable near-field radiative heat transfer via ferroelectric materials. *Applied Physics Letters*, *105*(24), 244102.
33. Amin, M., Farhat, M., & Bağcı, H. (2013). An ultra-broadband multilayered graphene absorber. *Optics express*, *21*(24), 29938-29948.
34. Zhang, Y., Feng, Y., Zhu, B., Zhao, J., & Jiang, T. (2014). Graphene based tunable metamaterial absorber and polarization modulation in terahertz frequency. *Optics express*, *22*(19), 22743-22752.
35. Xu, B. Z., Gu, C. Q., Li, Z., & Niu, Z. Y. (2013). A novel structure for tunable terahertz absorber based on graphene. *Optics express*, *21*(20), 23803-23811.
36. Zhihong, Z., Chucai, G., Jianfa, Z., Ken, L., Xiaodong, Y., & Shiqiao, Q. (2014). Broadband single-layered graphene absorber using periodic arrays of graphene ribbons with gradient width. *Applied Physics Express*, *8*(1), 015102.
37. RahimianOmam, Z., Ghobadi, A., Khalichi, B., & Ozbay, E. (2022). Adaptive thermal camouflage using sub-wavelength phase-change metasurfaces. Journal of Physics D: Applied Physics, 56(2), 025104.
38. Omam, Z. R., Ghobadi, A., Khalichi, B., & Ozbay, E. (2022). Fano resonance in a dolomite phase-change multilayer design for dynamically tunable omnidirectional monochromatic thermal emission. Optics Letters, 47(22), 5781-5784.
39. Gunes, F., Shin, H. J., Biswas, C., Han, G. H., Kim, E. S., Chae, S. J., ... & Lee, Y. H. (2010). Layer-by-layer doping of few-layer graphene film. ACS nano, 4(8), 4595-4600.
40. Kong, A., Cai, B., Shi, P., & Yuan, X. C. (2019). Ultra-broadband all-dielectric metamaterial thermal emitter for passive radiative cooling. *Optics express*, *27*(21), 30102-30115.
41. Zhao, B., Hu, M., Ao, X., Chen, N., & Pei, G. (2019). Radiative cooling: A review of fundamentals, materials, applications, and prospects. *Applied Energy*, *236*, 489-513.
42. Zhao, B., Hu, M., Ao, X., Chen, N., Xuan, Q., Jiao, D., & Pei, G. (2019). Performance analysis of a hybrid system combining photovoltaic and nighttime radiative cooling. *Applied energy*, *252*, 113432.
43. Dong, M., Chen, N., Zhao, X., Fan, S., & Chen, Z. (2019). Nighttime radiative cooling in hot and humid climates. *Optics express*, *27*(22), 31587-31598.
44. Rephaeli, E., Raman, A., & Fan, S. (2013). Ultrabroadband photonic structures to achieve high-performance daytime radiative cooling. *Nano letters*, *13*(4), 1457-1461.
45. Zhai, Y., Ma, Y., David, S. N., Zhao, D., Lou, R., Tan, G., ... & Yin, X. (2017). Scalable-manufactured randomized glass-polymer hybrid metamaterial for daytime radiative cooling. *Science*, *355*(6329), 1062-1066.
46. Granqvist, C. G., & Hjortsberg, A. (1981). Radiative cooling to low temperatures: General considerations and application to selectively emitting SiO films. *Journal of Applied Physics*, *52*(6), 4205-4220.
47. Eriksson, T. S., Jiang, S. J., & Granqvist, C. G. (1985). Surface coatings for radiative cooling applications: Silicon dioxide and silicon nitride made by reactive rf-sputtering. *Solar Energy Materials*, *12*(5), 319-325.
48. Ma, H., Yao, K., Dou, S., Xiao, M., Dai, M., Wang, L., ... & Zhan, Y. (2020). Multilayered $SiO_2/Si_3N_4$ photonic emitter to achieve high-performance all-day radiative cooling. *Solar Energy Materials and Solar Cells*, *212*, 110584.
49. Bartoli, B., Catalanotti, S., Coluzzi, B., Cuomo, V., Silvestrini, V., & Troise, G. (1977). Nocturnal and diurnal performances of selective radiators. *Applied Energy*, *3*(4), 267-286.
50. Hjortsberg, A., & Granqvist, C. G. (1981). Radiative cooling with selectively emitting ethylene gas. *Applied Physics Letters*, *39*(6), 507-509.
51. Hossain, M. M., Jia, B., & Gu, M. (2015). A metamaterial emitter for highly efficient radiative cooling. *Advanced Optical Materials*, *3*(8), 1047-1051.
52. Wang, X., Liu, X., Li, Z., Zhang, H., Yang, Z., Zhou, H., & Fan, T. (2020). Scalable flexible hybrid membranes with photonic structures for daytime radiative cooling. *Advanced Functional Materials*, *30*(5), 1907562.
53. Qu, Y., Li, Q., Du, K., Cai, L., Lu, J., & Qiu, M. (2017). Dynamic Thermal Emission Control Based on Ultrathin Plasmonic Metamaterials Including Phase-Changing Material GST. *Laser & Photonics Reviews*, *11*(5), 1700091.
54. Kocer, H., Cakir, M. C., Durna, Y., Soydan, M. C., Odabasi, O., Isik, H., ... & Ozbay, E. (2021). Exceptional adaptable MWIR thermal emission for ordinary objects covered with thin VO2 film. *Journal of Quantitative Spectroscopy and Radiative Transfer*, *262*, 107500.
55. Khalichi, B., Ghobadi, A., Osgouei, A. K., Kocer, H., & Ozbay, E. (2021). A Transparent All-Dielectric Multifunctional Nanoantenna Emitter Compatible With Thermal Infrared and Cooling Scenarios. *IEEE access*, *9*, 98590-98602.
56. Lumerical, F. D. T. D. (2018). Solutions, Inc.
57. Liu, X. L., & Zhang, Z. M. (2015). Giant enhancement of nanoscale thermal radiation based on hyperbolic graphene plasmons. Applied Physics Letters, 107(14), 143114.
58. Zhao, B., & Zhang, Z. M. (2015). Strong plasmonic coupling between graphene ribbon array and metal gratings. *ACS photonics*, *2*(11), 1611-1618.
59. He, M. J., Qi, H., Ren, Y. T., Zhao, Y. J., & Antezza, M. (2020). Magnetoplasmonic manipulation of nanoscale thermal radiation using twisted graphene gratings. *International Journal of Heat and Mass Transfer*, *150*, 119305.
60. He, M., Qi, H., Ren, Y., Zhao, Y., & Antezza, M. (2020). Active control of near-field radiative heat transfer by a graphene-gratings coating-twisting method. Optics letters, 45(10), 2914-2917.
61. He, M. J., Qi, H., Ren, Y. T., Zhao, Y. J., & Antezza, M. (2019). Graphene-based thermal repeater. *Applied Physics Letters*, *115*(26), 263101.
62. Ji, D., Cheney, A., Zhang, N., Song, H., Gao, J., Zeng, X., ... & Gan, Q. (2017). Efficient Mid-Infrared Light Confinement within Sub-5-nm Gaps for Extreme Field Enhancement. *Advanced Optical Materials*, *5*(17), 1700223.
63. Li, S. J., Wu, P. X., Xu, H. X., Zhou, Y. L., Cao, X. Y., Han, J. F., ... & Zhang, Z. (2018). Ultra-wideband and polarization-insensitive perfect absorber using multilayer metamaterials, lumped resistors, and strong coupling effects. *Nanoscale research letters*, *13*(1), 1-13.
64. Li, S. J., Li, Y. B., Li, H., Wang, Z. X., Zhang, C., Guo, Z. X., ... & Cui, T. J. (2020). A thin self-feeding Janus metasurface for manipulating incident waves and emitting radiation waves simultaneously. *Annalen der Physik*, *532*(5), 2000020.
65. Li, S. J., Li, Y. B., Zhang, L., Luo, Z. J., Han, B. W., Li, R. Q., ... & Cui, T. J. (2021). Programmable controls to scattering properties of a radiation array. Laser & photonics reviews, 15(2), 2000449.
66. Lide, D. R. (Ed.). (2004). *CRC handbook of chemistry and physics* (Vol. 85). CRC press.
67. Palik, E. D. (Ed.). (1998). *Handbook of optical constants of solids* (Vol. 3). Academic press.
68. Hanson, G. W. (2008). Dyadic Green's functions and guided surface waves for a surface conductivity model of graphene. *Journal of Applied Physics*, *103*(6), 064302.
69. Gómez-Díaz, J. S., & Perruisseau-Carrier, J. (2013). Graphene-based plasmonic switches at near infrared frequencies. *Optics express*, *21*(13), 15490-15504.
70. Amin, M., Farhat, M., & Bağcı, H. (2013). A dynamically reconfigurable Fano metamaterial through graphene tuning for switching and sensing applications. Scientific reports, 3(1), 1-8.
71. Huang, Z., & Ruan, X. (2017). Nanoparticle embedded double-layer coating for daytime radiative cooling. *International Journal of Heat and Mass Transfer*, *104*, 890-896.
72. Berk, A., Conforti, P., Kennett, R., Perkins, T., Hawes, F., & Van Den Bosch, J. (2014, June). MODTRAN® 6: A major upgrade of the MODTRAN® radiative transfer code. In *2014 6th Workshop on Hyperspectral Image and Signal Processing: Evolution in Remote Sensing (WHISPERS)* (pp. 1-4). IEEE.
73. Li, J., Gan, R., Guo, Q., Liu, H., Xu, J., & Yi, F. (2018). Tailoring optical responses of infrared plasmonic metamaterial absorbers by optical phonons. *Optics express*, *26*(13), 16769-16781.
74. Palik, E. D., & Ghosh, G. C. (1997). Handbook of thermo-optic coefficients of optical materials with applications.
75. Rossi, E., Adam, S., & Sarma, S. D. (2009). Effective medium theory for disordered two-dimensional graphene. *Physical Review B*, *79*(24), 245423.
76. Ghobadi, A., Hajian, H., Butun, B., & Ozbay, E. (2018). Strong light–matter interaction in lithography-free planar metamaterial perfect absorbers. *ACS Photonics*, *5*(11), 4203-4221.
77. Nielsen, M. G., Gramotnev, D. K., Pors, A., Albrektsen, O., & Bozhevolnyi, S. I. (2011). Continuous layer gap plasmon resonators. Optics express, 19(20), 19310-19322.
78. Nielsen, M. G., Pors, A., Albrektsen, O., & Bozhevolnyi, S. I. (2012). Efficient absorption of visible radiation by gap plasmon resonators. Optics express, 20(12), 13311-13319.
79. Ding, F., Yang, Y., Deshpande, R. A., & Bozhevolnyi, S. I. (2018). A review of gap-surface plasmon metasurfaces: fundamentals and applications. *Nanophotonics*, *7*(6), 1129-1156.
80. Pan, M., Huang, Y., Li, Q., Luo, H., Zhu, H., Kaur, S., & Qiu, M. (2020). Multi-band middle-infrared-compatible camouflage with thermal management via simple photonic structures. *Nano Energy*, *69*, 104449.
81. Zhu, H., Li, Q., Tao, C., Hong, Y., Xu, Z., Shen, W., ... & Qiu, M. (2021). Multispectral camouflage for infrared, visible, lasers and microwave with radiative cooling. *Nature communications*, *12*(1), 1-8.